\documentclass[iop]{emulateapj}
\usepackage{apjfonts}

\newcommand{\be}{\begin{equation}}
\newcommand{\ee}{\end{equation}} 
\newcommand{\ba}{\begin{eqnarray}}
\newcommand{\ea}{\end{eqnarray}}

\begin{document}

\epsscale{1.1}

\title{Probing Reionization with Intensity Mapping of Molecular and Fine Structure Lines }
\shorttitle{}

\author{
Yan Gong\altaffilmark{1}, 
Asantha Cooray\altaffilmark{1}, 
Marta B. Silva\altaffilmark{2},
Mario G. Santos\altaffilmark{2},
Phillip Lubin\altaffilmark{3}
}

\altaffiltext{1}{Department of Physics \& Astronomy, University of California, Irvine, CA 92697}
\altaffiltext{2}{CENTRA, Instituto Superior T\'eecnico, Lisboa 1049-001, Portugal}
\altaffiltext{3}{Department of Physics \& Astronomy, University of California, Santa Barbara, CA}

%\submitted{}

\begin{abstract}
We propose observations of the molecular gas distribution during the era of reionization.  At $z \sim $ 6 to 8, $^{12}$CO$(J=1-0)$ line intensity results
in a mean brightness temperature of about 0.5 $\mu$K with rms fluctuations of  0.1 $\mu$K at 1 to 10 Mpc spatial scales,
corresponding to 30 arcminute angular scales. This intensity fluctuations
can be mapped with an interferometer, similar to existing and planned 21-cm background experiments, but  operating at $\sim$ 12 to 17 GHz.
We discuss the feasibility of detecting the cross-correlation between HI and CO molecular gas since such a cross-correlation has the advantage that
it will be independent of systematics and most foregrounds in each of the 21-cm and CO$(1-0)$ line experiments. Additional instruments tuned to higher-order
transitions of the CO molecule or an instrument operating with high spectral resolution at mm-wavelengths targeting 158 $\mu$m CII 
could further improve the reionization studies with molecular gas. The combined 21-cm and CO line observations has the potential
to establish the relative distribution of gas in the inter-galactic medium and molecular gas that are clumped in individual first-light 
galaxies that are closely connected to the formation of massive stars in these galaxies.
\end{abstract}


\section{Introduction} 
Recently there has been a great deal of interest in the evolution of the intergalactic medium at high redshifts ($z > 6$) 
and specifically, the reionization history of the Universe. 
Most of this interest is sparked by existing multi-wavelength observations that suggest a complex reionization history, with a strong possibility for
 extended and highly inhomogeneous period of reionization. 
Observations of the 21-cm spin-flip line of neutral hydrogen are 
currently considered to be one of the most promising probes of the epoch of reionization 
(e.g., Madau et al. 1997; Loeb \& Zaldarriaga 2004; Gnedin \& Shaver 2004).
Given the line emission, leading to a frequency selection for observations, the 21 cm data provides a tomographic view of the reionization 
(Santos et al. 2005; Furlanetto et al. 2004a). It is also a useful cosmological probe 
(Mao et al. 2008; Santos \& Cooray 2006; McQuinn et al. 2006; Bowman et al. 2007).

Here we propose spectral line intensity mapping associated with molecular gas during the era of reionization as an additional probe
during the formation of first-light galaxies. In particular, we study the $J=1$ to 0 transition of the $^{12}$CO (carbon monoxide),
which has a rest-wavelength of 2.61 mm. Our suggestion extends the initial work on this topic by Righi et al. (2008; see also Basu et al. 2004).
Instead of the angular power spectrum, which is dominated by CMB anisotropies at tens of arcminute scales, we propose spectral line intensity mapping 
leading to the three-dimensional power spectrum (see also Visbal \& Loeb 2010).  An additional probe of reionization is useful since
the first-generation 21-cm experiments are expected to be noise-dominated and the observations could, in principle, be affected by various
systematics and residual foregrounds.  A second probe of the reionization epoch  providing three-dimensional information can be
 cross-correlated with the 21-cm spectral line measurements to improve the overall understanding of reionization. 

In terms of potential spectral lines of interest, the 
line intensity of the $J=1$ to 0 transition of $^{12}$CO has been used as a way to establish the gas mass of star-forming galaxies both at low 
(e.g., Downes \& Solomon 1998) and medium (e.g., Greve et al. 2005) redshifts. Individual CO 
line observations now exist out to $z=5.3$ (e.g., Riechers et al. 2010), which 
is significant given that the highest 21-cm line emission from a galaxy currently
measured is at $z$ of 0.24 (Lah et al. 2007; see Chang et al. 2010 for 21-cm intensity variations
of the kind we propose 
at $z\sim 0.9$).  In addition to the CO lines, we also find that the fine structure CII line, with observations in the mm range of
the spectrum, could be a strong probe of reionizaion. 

This {\it Letter} is organized as follows: in the next Section, we outline the calculation related to the strength of the expected signal from reionization, Section~3 presents results related to a potential CO brightness temperature fluctuation experiment that target 1-0 transition,
and Section~4 discusses the cross-correlation with the 21-cm background.

\section{The CO(1-0) signal}

\begin{figure*}[htbp]
\centerline{
\includegraphics[scale = 0.45]{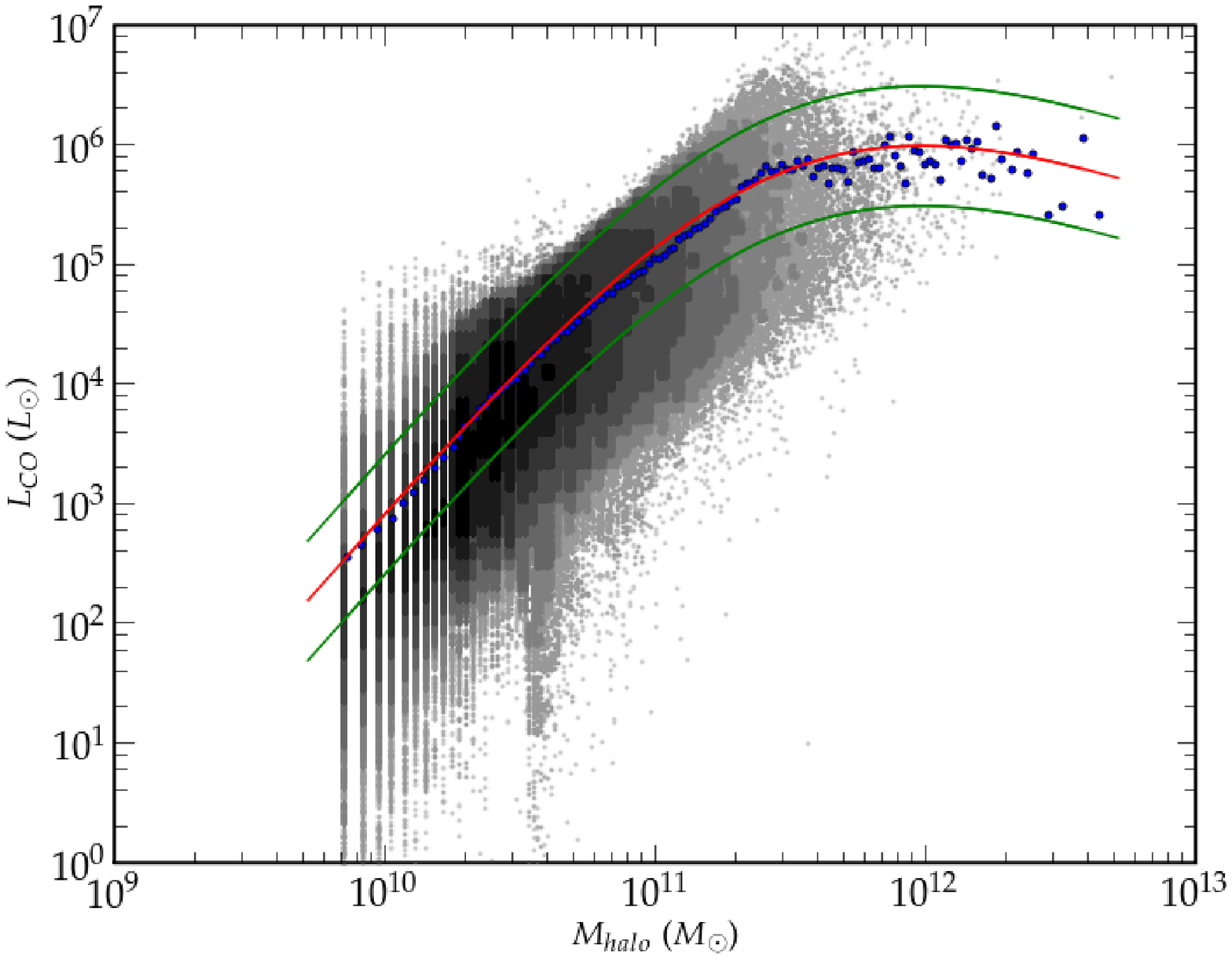}
\includegraphics[scale = 0.45]{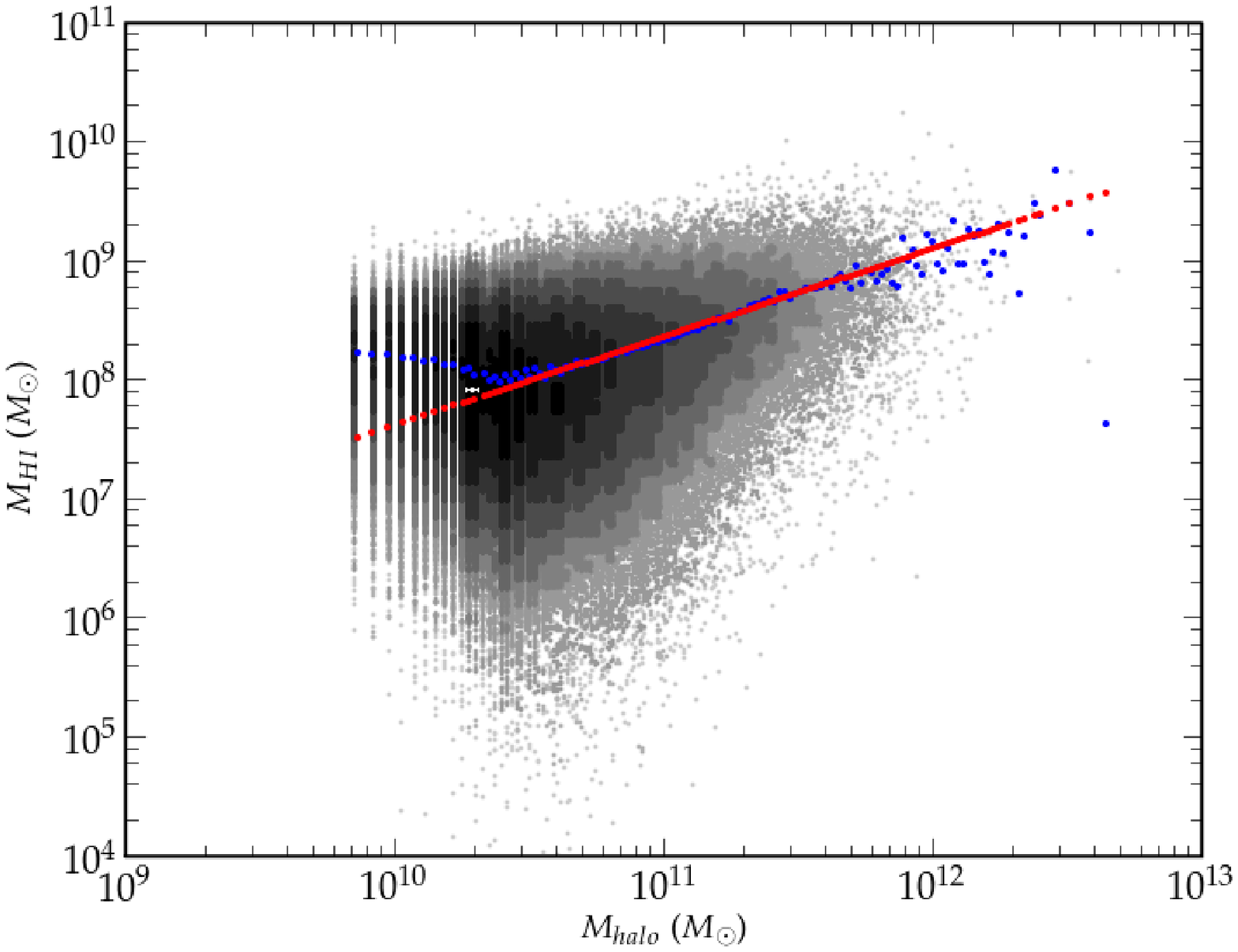} 
}
\caption{\label{fig:L_CO_M_HI}  
{\it Left}: The $L_{\rm CO}(M_{\rm halo})$ relation for CO(1-0) luminosity as a function of the halo mass $M_{\rm halo}$ at $z=7$. The lines shows the
mean relation (thick center line) and $\pm 1\sigma$ relation (thin lines) given the scatter (see text for details).
The dots show the mean of the scatter when binned to 150 logarithmic intervals in halo mass.
{\it Right:} The relation  neutral hydrogen mass content in individual galaxies $M_{\rm HI}$ vs. the halo mass
 $M_{\rm halo}$ at $z=7$.}
\end{figure*}

For a galaxy at a given redshift the CO(1-0) luminosity is taken to be related to the ${\rm H}_2$ content
and the metallicity of the gas (Obreschkow et al. 2009b). A key input here is the
relation between CO(1-0) luminosity, $L_{\rm CO}$, and the halo mass. While such a relation is expected it also has a large scatter
coming from differences in the gas content, at a fixed halo mass, due to variations in the halo ages at a fixed mass;
The gas in older halos is more likely to be depleted more than the younger ones with older halos containing more stars.
However, for sufficiently large volumes,  we will be averaging over many galaxies so that, so the the mean relation and its scatter can be quantified.
This is similar to the luminosity-halo mass relation for optical light which has allowed statistics 
such as the conditional luminosity functions to be constructed and improve the halo modeling beyond the simple halo occupation number
(e.g., Yang et al. 2004; Cooray \& Milosavljevic 2005b; Cooray 2006).

Here we make use of the numerical simulations of Obreschkow et al. (2009d) that are 
available as part of the SKA Simulated Skies website\footnote{http://s-cubed.physics.ox.ac.uk}.
This simulation is based on the galaxy catalog from De Lucia $\&$ Blaizot (2007) which post-processed the Millennium dark matter simulation with semi-analytical models of galaxy formation.
It captures the gas astrophysics, especially HI and H$_2$
and allow for the cosmic evolution of these two phases of the cold hydrogen gas (see, Obreschkow et al. 2009b; 2009c for details).
They also constructed a virtual sky field with HI and CO line luminosities for each of the galaxies, calculated from a combination of the neutral
and molecular hydrogen gas masses and the metallicity of each galaxy. Using the online tools we made a query with a four degree observing cone 
containing more than $5\times 10^5$  galaxies at each of $z=7$.

The $L_{\rm CO}(M)$ from every halo at $z \sim 7$ from this simulation
is shown in the left panel of Figure~\ref{fig:L_CO_M_HI} with the shade of gray darker in the regions where the halo density is higher. 
 Typically we find one galaxy per halo at these redshifts.
The lines show the result of averaging the total $L_{\rm CO}$ from the halos for each of 150 logarithmic bins of halo mass.  
Motivated by parameterization of optical luminosity and halo mass (Cooray \& Milosavljevic 2005a),
we use a relation of the form $L_{\rm CO}(M) = L_0(M/M_c)^b(1+M/M_c)^{-d}$ 
to describe the mean relation and determine the four free parameters $A$, $b$, $c$ and $d$.
At $z=6,7,$ and 8, these parameters take the values
of $L_0=4.3\times 10^{6},6.2\times 10^{6}, 4.0\times 10^{6}$ L$_{\sun}$, 
$b=2.4,2.6,2.8$, $M_c= 3.5\times 10^{11}, 3.0\times 10^{11}, 2.0\times 10^{11}$ M$_{\sun}$, 
and $d=2.8,3.4,3.3$, respectively.

\begin{figure}[htbp]
\includegraphics[scale = 0.45]{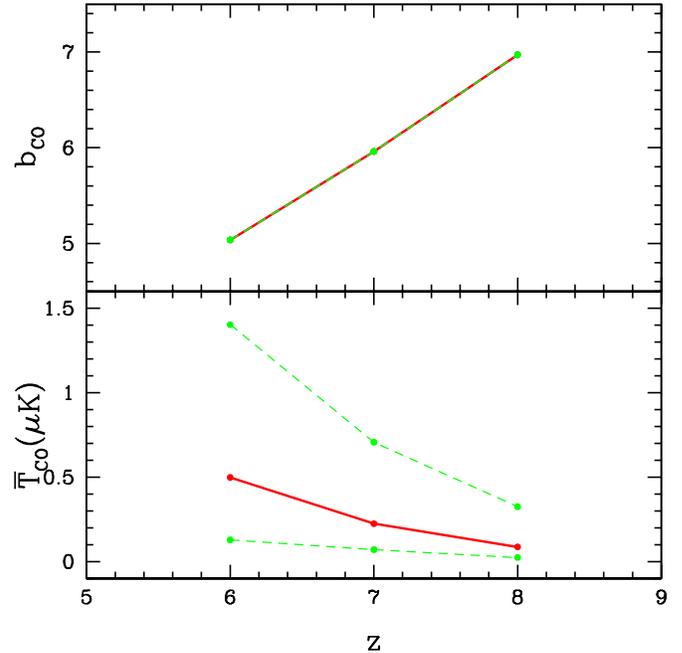} 
\caption{\label{fig:biasTco} The bias and brightness 
temperature of CO line for $z=6$, $7$ and $8$. The $68.3\%$
C.L. ($1\sigma$) are also shown in green dashed lines. The bias has no additional uncertainty as the uncertainty in $L_{\rm CO}(M)$ relation is
a change in the overall amplitude at a given mass and not a shape change (see. eq. 2).}
\end{figure}

\begin{figure*}[htbp]
\centerline{
\includegraphics[scale = 0.27]{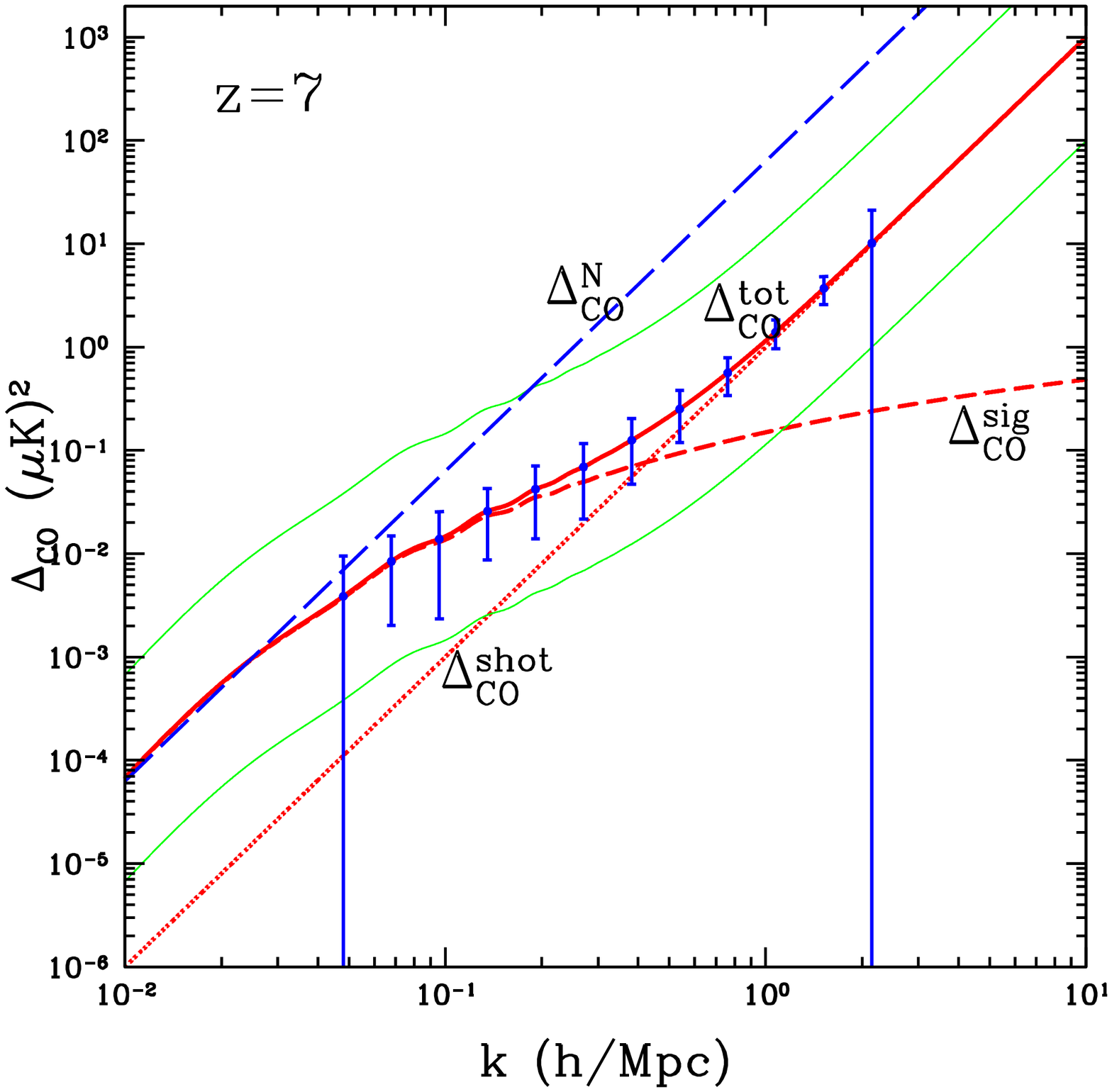}
\includegraphics[scale = 0.27]{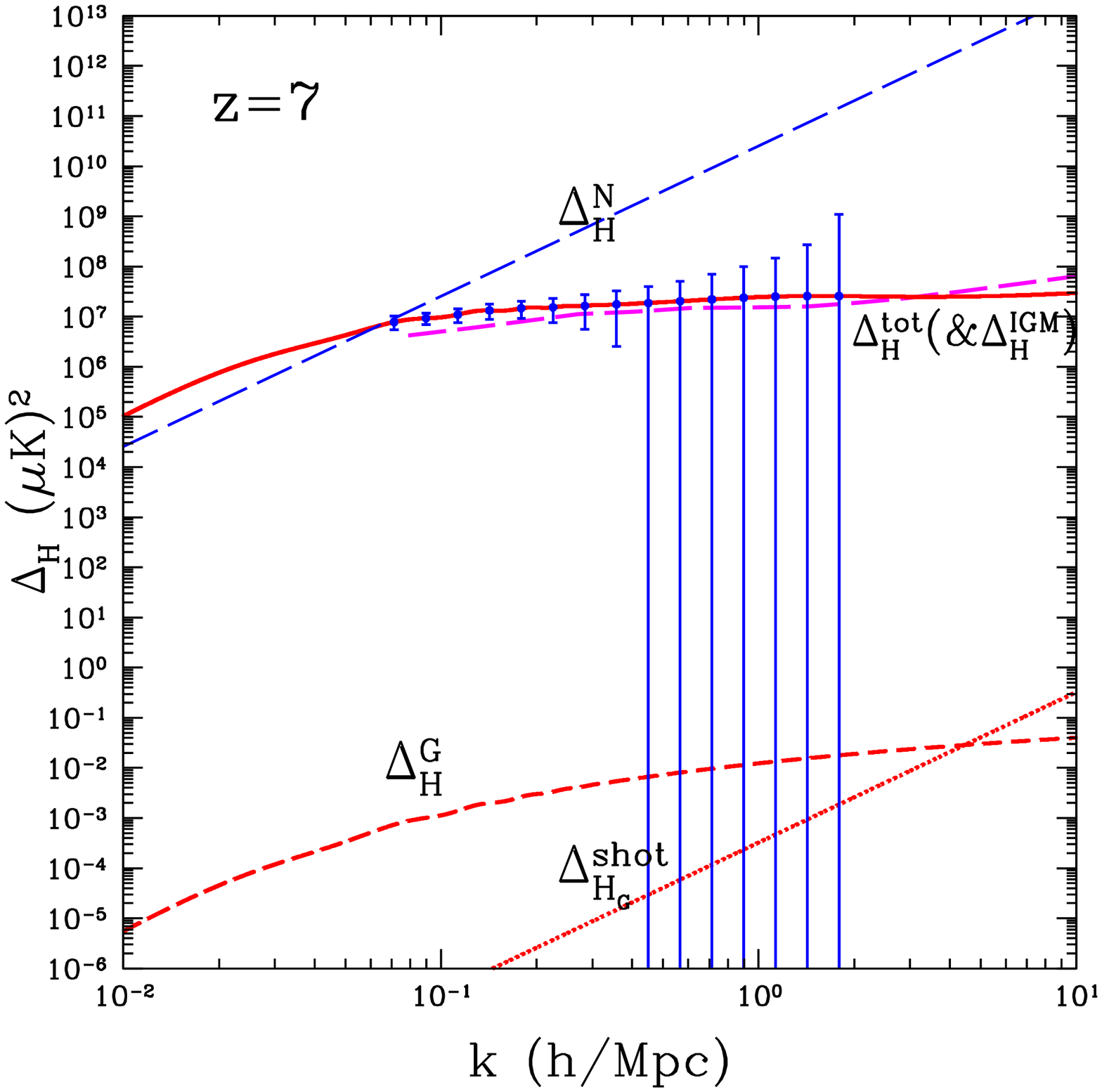} 
\includegraphics[scale = 0.27]{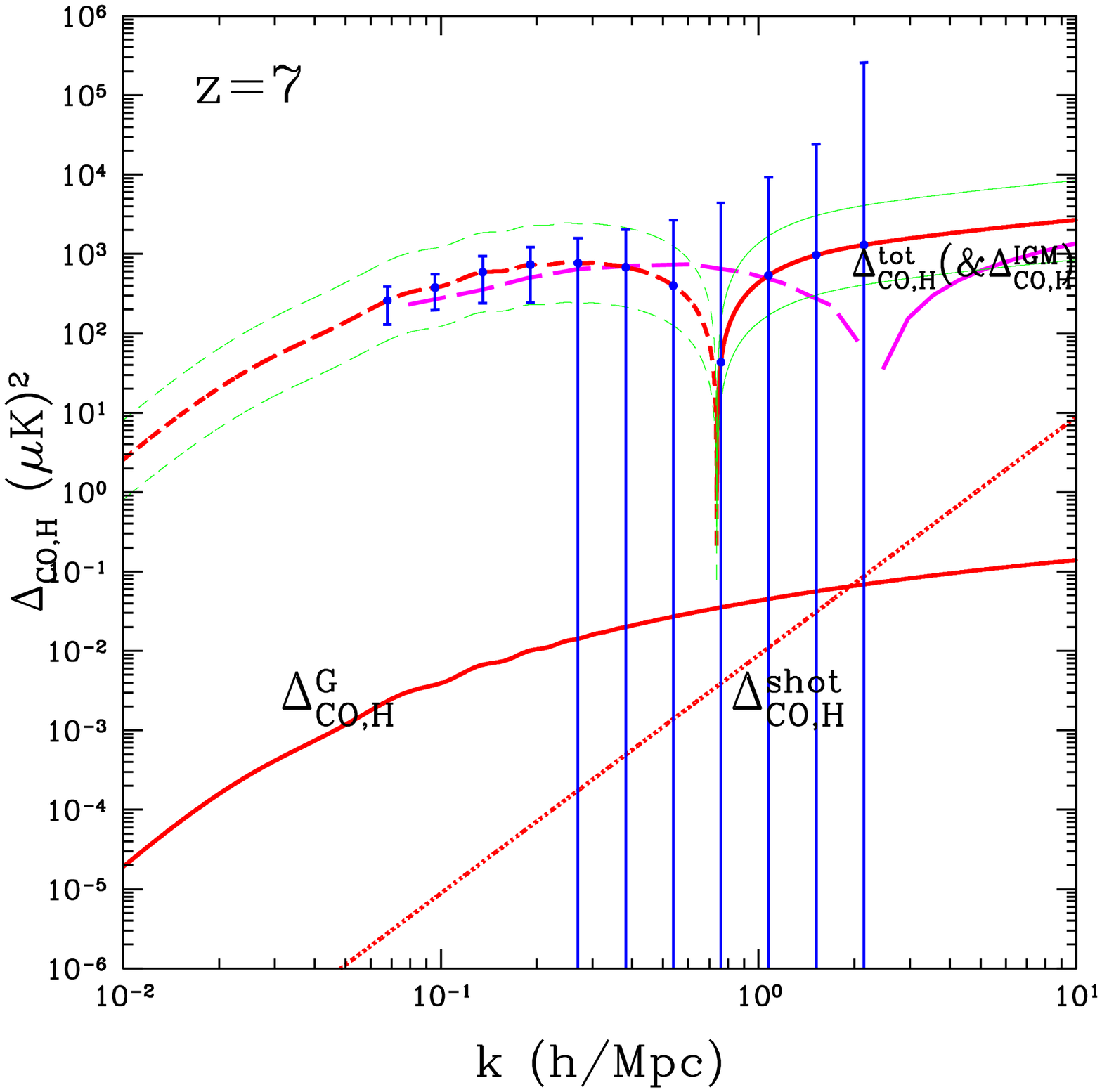} 
}
\caption{\label{fig:Pco} Left panel: the dimensionless auto power 
spectrum and the shot noise power spectrum for CO line
at $z=7$. The $68.3\%$ C.L. ($1\sigma$) is shown in green thin lines.
Middle panel: the dimensionless auto power 
spectrum and the shot noise power spectrum 
for 21-cm emission at $z=7$. The auto and shot noise
power spectrum of the galaxies $\Delta_{\rm H}^{\rm G}$
and $\Delta^{\rm shot}_{\rm H_G}$ are so small compared
with $\Delta_{\rm H}^{\rm IGM}$, that 
$\Delta^{\rm tot}_{\rm H}$ is dominated by 
$\Delta_{\rm H}^{\rm IGM}$.
The blue lines with error bars make use of noise power spectrum (shown by the blue dashed line) corresponding to the design of LOFAR.
Right panel: the dimensionless cross power 
spectrum and shot noise power spectrum for 
CO line and 21-cm emission at $z=7$. The blue solid 
line is the error from LOFAR.
The $68.3\%$ C.L. ($1\sigma$) is shown in red thin lines.
The long-dashed magenta lines in the middle and right panels
are the results from the simulation of 21-cm in Santos
et al. (2010) for comparison.}
\end{figure*}

Using the above relation between $L_{\rm CO}$ and halo mass $M$ as a function of redshift, we can write the 
mean intensity of the CO(1-0) line as
\be
\bar{I}_{\rm CO} = \int_{M_{\rm min}}^{\infty} dM \frac{dn}{dM}(z,M)\frac{L_{\rm CO}(z,M)}{4\pi D_{\rm L}^2}y(z)D_{\rm A}^2,
\label{eq:I_ave}
\ee
where we take $M_{\rm min} = 10^8 (M_{\odot}/h)$ as the minimum 
mass of the dark matter halos that can host galaxies (to be consistent with mass scale of atomic Hydrogen cooling or a virial temperature of 10$^4$K 
(Loeb \& Barkana 2001), $dn/dM$ is the halo mass function (Sheth \& Tormen 1999), 
$D_{L}$ is the 
luminosity distance, and $D_{\rm A}$ is the comoving angular 
diameter distance. In above $y(z)={d\chi}/{d\nu}={\lambda_{\rm CO}(1+z)^2}/{H(z)}$, where
 $\chi$ is the comoving distance, $\nu$ is the observed
frequency, $\lambda_{\rm CO}=2.6$ mm is the rest frame  wavelength of the CO(1-0) line. 

The signal is expected to contain spatial variations around the average intensity due to poisson fluctuations in the number of halos (shot noise) and correlations with the underlying dark matter density field which will enhance the fluctuations by a factor of $(1+b(z,M)\delta)$, with $\delta$ being the density contrast of this dark matter field.
The intensity of the CO(1-0) line can be written as a function of the spatial location 
$I_{\rm CO}({\bf x}) = \bar{I}_{\rm CO}[1+b_{\rm CO}\delta({\bf x})]$ with bias
$b_{\rm CO}$ given by
\be
b_{\rm CO}(z)=\frac{\int_{M_{\rm min}}^\infty dM\frac{dn}{dM}L_{\rm CO}b(z,M)}
{\int_{M_{\rm min}}^\infty dM\frac{dn}{dM}L_{\rm CO}},
\ee
where $b(z,M)$ is the halo bias (Sheth \& Tormen 1999).

In Figure~2, we show both the mean signal and the halo bias  as a function of the redshift during reionization. For simplicity,
we have converted the mean intensity of the signal to Rayleigh-Jeans brightness temperature. At $ z\sim 6$,
the signal is at the level of 0.2 to 1.5 $\mu$K, with the large range coming from the scatter in the $L_{\rm CO}(M)$ relation.

Since the mean intensity is challenging to measure as it requires absolute measurements
we discuss the feasibility of measuring anisotropies of the CO line intensity.
The total power spectrum of the CO(1-0) line involves two contributions, the clustering term and the shot noise.
Following the mean intensity and bias given above, the clustering power spectrum is 
\be
P_{\rm CO}^{\rm clus}(z,k)=\bar{T}_{\rm CO}^2 b_{\rm CO}^2 P_{\delta\delta}(z,k).
\ee
Here $\bar{T}_{\rm CO}$ is the mean temperature of CO line,
$P_{\delta\delta}=P_{\rm lin}(z,k)$ where $P_{\rm lin}$ is the linear
power spectrum of the dark matter. The shot niose power spectrum, due to 
discretization of the galaxies, is
\be
P^{\rm shot}_{\rm CO}(z) = \int_{M_{\rm min}}^{\infty} dM \frac{dn}{dM} 
\bigg(\frac{L_{\rm CO}}{4\pi D_{\rm L}^2}yD_{\rm A}^2\bigg)^2.
\ee

\section{Cross power spectrum of CO line and 21-cm emission}

In order to calculate the expected strength of the correlation between 21-cm and CO line intensities, we briefly discuss the
brightness temperature variations in the 21-cm signal. We consider two signals, the usual signal coming from neutral gas in the
intergalactic medium that is not directly correlated with molecular gas in galaxies, and the remaining neutral gas in individual galaxies.
During the era of reionization the 21-cm brightness temperature fluctuations are expected to be dominated by the IGM.
At lower redshifts, While the IGM signal is zero, a small 21-cm
contribution still remains due to the neutral gas in individual galaxies.

The difference of the spatially averaged brightness 
temperature of 21-cm emission in the IGM and the CMB temperature at redshift 
z (Mao et al. 2008) is
\be
\bar{T}^{\rm IGM}_{\rm b}=c(z)\bar{x}_{\rm H}\bigg(\frac{\bar{T}_{\rm s}-T_{\rm CMB}}{\bar{T}_{\rm s}}\bigg)\ (\rm mK).
\ee
Here $\bar{x}_{\rm H}$ is the mean neutral hydrogen fraction,
that we assume $\bar{x}_{\rm H}\equiv 1-\bar{x}_{\rm i}$
where $\bar{x}_{\rm i}$ is the mean ionized fraction, and
we set $\bar{x}_{\rm H} = 0.3$ and $0.7$ at $z = 7$ and $8$
respectively (Mao et al. 2008). 
The $\bar{T}_{\rm s}$ is the averaged spin temperature of the
IGM, and we assume $\bar{T}_{\rm s}\gg T_{\rm CMB}$ here.
The parameter $c(z)$ (defined in Santos et al. 2008) is
\be
c(z) = 23\bigg(\frac{0.7}{h}\bigg)\bigg(\frac{\Omega_{\rm b}h^2}{0.02}\bigg)
\bigg(\frac{0.15}{\Omega_{\rm m}h^2}\frac{1+z}{10}\bigg)^{1/2} ({\rm mK}),
\ee
where $\Omega_{\rm b}$ is the baryon density parameter.

We obtain the power spectrum of the 21-cm fluctuations in the IGM using the fitting formula
of Mao et al. (2008) at several redshifts during reionization.
For comparison we also consider the reionization simulations of Santos et al. (2010).
These simulations have a smaller bubble size on average than the models of
Mao et al. (2008). As we will discuss later, while two models we consider give rise to the same 21-cm
power spectrum, they different significantly in terms of the cross-correlation of CO and 21-cm
brightness temperature fluctuations.

Similar to the IGM calculation the mean brightness temperature of 21-cm emission from the
galaxies can be calculated through
\be
\bar{T}^{\rm G}_{\rm b} = c(z)\frac{\rho_{\rm H}(z)}{X\rho_{\rm b}(z)}\, (\rm mK),
\ee
where $X=0.76$
is the hydrogen mass fraction, $\rho_{\rm b}(z)=\Omega_{\rm b}
(1+z)^3\rho_{\rm c}$ is the baryon density at z and 
$\rho_{\rm c}= 2.7752 \times 10^{11}\ {\rm M_{\odot}}/h$ $({\rm Mpc}/h)^{-3}$. 

We assume that the total hydrogen mass in a given pixel of the experiment only depends on the halo mass
so that the hydrogen mass density $\rho_{\rm H}$ is given by
$\rho_{\rm H} = \int_{M_{\rm min}}^{\infty} dM \frac{dn}{dM}M_{\rm HI}(M)$.
To obtain $M_{\rm HI}(M)$ we used the same procedure as before, querying the simulation from Obreschkow et al. (2009d) and adding the $M_{\rm HI}$ from the galaxies in each halo. 
The resulting scatter plot is shown in the right panel of Figure~\ref{fig:L_CO_M_HI} together with the average in each mass bin (blue dots). The red 
line shows a fitting function to the average values using
$M_{\rm HI}=10^{A\times\log_{10}(M)+B}$,
where $M$ is the halo mass, $A=0.75$, $0.74$ and $0.75$,
and $B=0.17$, $0.23$ and $0.07$ for $z=6$, $7$ and $8$
respectively.

Following the description related to the CO line intensity, the mean 21-cm
brightness temperature is taken to have spatial variations
$T^{\rm G}_{\rm b}({\bf x}) = \bar{T}^{\rm G}_{\rm b}(1+b_{\rm H}\delta({\bf x}))$,
where $b_{\rm H}$ is the bias of galaxies containing neutral hydrogen 
\be
b_{\rm H}(z) = \frac{\int_{M_{\rm min}}^{\infty} dM \frac{dn}{dM} M_{\rm H} b(z,M)}{\rho_{\rm H}}.
\ee
The clustering power spectrum of the 21-cm emission from galaxies is then $P^{\rm G}_{\rm H}(z,k) = (\bar{T}^{\rm G}_{\rm b})^2 b^2_{\rm H}P_{\delta \delta}(z,k)$.
Again, the shot noise power spectrum for 21-cm emission is
\be
P^{\rm shot}_{\rm H}(z) = \int_{M_{\rm min}}^{\infty} dM \frac{dn}{dM}
\bigg(c(z)\frac{M_{\rm H}}{X\rho_{\rm b}(z)} \bigg)^2.
\ee
The total power spectrum of 21-cm emission follows from a sum of shot-noise and clustering terms 
(see middle panel in figure~\ref{fig:Pco}).

The cross power spectrum between 21-cm and CO line intensities is 
$P^{\rm tot}_{\rm CO,H}\ =\ P^{\rm IGM}_{\rm CO,H} + P^{\rm G}_{\rm CO,H} + P^{\rm shot}_{\rm CO,H}$,
where we again separate the contribution from neutral hydrogen in the IGM and neutral hydrogen remaining in individual galaxies.
Based on the discussion related to 21-cm fluctuations, the cross power spectrum of the CO line 
and 21-cm emission in the IGM is
\be
P^{\rm IGM}_{\rm CO,H}(z,k) = \bar{T}_{\rm CO}b_{\rm CO}c(z)
[(1-\bar{x}_{\rm i})P_{\delta\delta}-\bar{x}_{\rm i}P_{\delta_{x}\delta}],
\ee
where $P_{\delta_{x}\delta}(z,k)$ is the cross power 
spectrum for the ionized fraction and the dark matter.
The IGM gas is correlated with CO lines as they both trace the same underlying dark matter density field.

The cross power spectrum of the CO line and 21-cm emission in individual galaxies is
\be
P^{\rm G}_{\rm CO,H}(z,k) = \bar{T}_{\rm CO}\bar{T}^{\rm G}_{\rm b}b_{\rm CO}b_{\rm H}P_{\delta\delta}(z,k),
\ee
while the shot noise power spectrum of the CO line and 21-cm emission
in the galaxies is
\be
P^{\rm shot}_{\rm CO,H}(z) = \int_{M_{\rm min}}^{\infty} dM \frac{dn}{dM}
\frac{L_{\rm CO}}{4\pi D_{\rm L}^2}yD_{\rm A}^2 c(z) \frac{M_{\rm H}}{X\rho_{\rm b}(z)}.
\ee

\section{Discussion}

In Figure~3, left panel, we show the brightness temperature power spectrum for the CO line at $z=7$, while in
the middle panel we show the 21-cm brightness temperature fluctuations at the same redshift and in the right panel, 
we show the cross-correlation between the two. In the case of 21-cm fluctuations, the solid lines show the prediction based on the models of Mao et al.
(2008), while in dashed lines we show the prediction from the semi-numerical simulation of Santos et al. (2010).

To discuss the possibility of measuring the CO signal and the CO-21 cm cross-correlation
we make use of a calculation similar to the one used in 21-cm interferometers.
Following Santos et al. (2010), the noise power spectrum, $P_N$ for a given experimental ``pixel'' in Fourier space is given by
$P_N(k,\theta)=D_A^2 y (\lambda^4 T_{sys}^2)/[A_e^2 t_0  n\left(D_A k \sin(\theta)/2\pi\right)]$
where  $A_e$ is the collecting area of one element of the interferometer (which could be a station), $t_0$ is the total observation time and the function $n()$ captures the 
baseline density distribution on the plane perpendicular to the line of sight, assuming already it is 
rotationally invariant ($k$ is the moduli of the wave mode $\mathbf{k}$ and $\theta$ is the angle between $\mathbf{k}$ and the line of sight). 
Typically this baseline distribution will be more concentrated around the core but for interferometers with high filling factors we can make the simplification that the distribution is constant so that the required function for above is $n()=\lambda^2 N_a^2/\pi D_{max}^2$ with
$N_a$ as the number of elements of the interferometer and $D_{max}$ the maximum baseline of the distribution. Note that the error in the measured power spectrum 
at a given $\mathbf{k}$ is just $\left(P_S+P_N\right)/\sqrt{(N_m)}$, where $P_S$ is the signal power spectrum
and $N_m$ is the number of modes contributing to a given binned measurement. In order to count these number of modes, we consider a grid in $k$ space with 
a resolution given by $(2\pi)^3/({\rm y r^2\cdot {\rm FoV}\cdot B})$, which is set by the volume of the experiment ($B$ is the bandwidth used in the analysis and FoV is the field of view).

Typically we will be interested on linear scales ($k<1.0$ h/Mpc) corresponding to a few arcminutes. 
Given the frequencies involved, the need for large FoV, and high sensitivity to $\mu$K brightness temperature fluctuations leads to
an experimental setup that involves a total collecting area $A_{\rm tot} = 385$ m$^2$, bandwidth $B=1$ GHz with
spectral resolution $\delta \nu=30$ MHz and interferometer spacings between 0.7 and 25 m. The number of elements is
1000 with each having a receiver with system temperature of 20K. In the left panel of Figure~3, we assume a total integration time of 3000 hours when calculating the noise power spectrum.
For 21-cm noise calculations we make use of a setup similar to LOFAR, though we note that our calculations are not too different for
an experiment like MWA. For experimental parameters related to LOFAR, we refer the reader to Harker et al. (2010). The values (at 150 MHz) are
$A_{\rm tot}=5\times10^4$ m$^2$, $T_{\rm sys}=490$K (instrument noise temperature), 
FOV=25 deg$^2$, bandwidth of 8 MHz with $\delta \nu=0.5$ MHz and $D_{\rm max}=2000$ m. We take $T_{\rm int}=1000$ hours.

For the proposed experimental setup the brightness temperature fluctuations can be detected with a cumulative signal-to-noise ratio that leads to
more than 40$\sigma$ confidence. However, CO(1-0) mapping could be impacted by foregrounds, including spectral lines associated with
other molecules; In the case of other transitions of the CO lines, the contamination for CO(1-0) measurements at $z\sim$ 6 to 8 
is likely to be minimal; the CO(2-1) line for example must originate from $z \sim$ 13 to 17. The most significant contaminations will
likely come from molecular transitions that have rest wavelengths longward of the CO(1-0) line and thus coming
from lower redshifts. To avoid such contaminations and to improve the overall study of reionization  we suggest a
cross-correlation between 21-cm brightness temperature fluctuations and the CO intensities.

For the cross-correlation, we assume that observations will be conducted in overlapping areas on the sky by both the CO and 21-cm experiment.
This is a necessary feature of cross-correlation study leading to some coordination between two experiments.
We reduce both datasets to a common, and a lower resolution cube in frequency and count the total number of common modes contributing to a given bin in $k$ 
space using the same method for noise calculation as before for each of the experiments.
Note that the error in a given pixel in $k$ space can be written as $\sqrt{(P_{\rm CO,H}^2+P_{\rm CO}P_{\rm H})/N_m}$, where $N_m$ is the number of modes.
In Figure~3 right label we show the cross-correlation power spectrum at $z=7$ and the expected binned errors by using LOFAR and the CO experiment with parameters
outlined in above. The overall uncertainties are dominated by the 21-cm observations. For reference, we also show the prediction of the model by Mao et al. (2008)
and from  the numerical simulation of Santos et al. (2010). While the two give a similar 21-cm fluctuation power spectrum, they differ significantly in terms 
of the cross-correlation with the CO emission. The difference is primarily related to the average scale of the reionized bubble sizes and
we find that the cross-correlation such as the one we propose here is more sensitive to astrophysics during reionization than with 21-cm data alone.

While our calculation and results are related to the CO(1-0) transition, our method can be extended to other transitions and other molecules as well as long
as a correction is made for the ratio of line intensity between higher J transitions of CO and the 1-0 line.
Existing observations, at low to moderate redshifts, suggest a value of about 0.6 for $J=2$ to 1 luminosity when compared to 1-0 luminosity.
The 2-1 observations will be over the frequency range of 25.5 to 32 GHz for $z \sim 6$ to 8.

The fine structure line CII has been detected in several of the high-z galaxies and existing studies indicate that $L_{\rm CII}/L_{\rm CO} \sim 4100$.
With a rest wavelength at 158 $\mu$m, observations are required over the frequency range of 210 to 270 GHz (1.1 to 1.4mm).
The rms fluctuations are expected to be at the level of $10^2$ to 10$^3$ Jy/sr at $z \sim 7$ to 8 at tens of arcminute angular scales.
To probe reionization fluctuations associated with the CII line,  observations must be carried out in small frequency intervals of 50 MHz with instruments providing
angular resolution of order  ten arcminutes with survey areas of order a few tens square degrees. 
In a future paper we hope to return to the scientific case of a  potential CII experiment, since such an observation can probe multiple redshifts with both
CII and high-J CO lines.

\acknowledgements
We thank participants of the Kavli Institute for Space Studies' (KISS) Billion Years workshop for helpful discussions. This work was
supported by NSF CAREER AST-0645427 at UCI. MGS and MBS acknowledges support from FCT-Portugal under grant PTDC/FIS/100170/2008.

\end{document}